\documentstyle[prb,aps]{revtex}
\begin{document}
\draft

\title{Low-temperature behavior of the large-{\boldmath $\!\!U$} Hubbard model
\\ from high-temperature expansions}

\author{D.F.B. ten Haaf, P.W. Brouwer, P.J.H. Denteneer, and J.M.J. van
Leeuwen}
\address{Institute Lorentz, Leiden University,
P.O. Box 9506, 2300 RA Leiden, The Netherlands}

\date{July 11, 1994}

\maketitle

\begin{abstract}
We derive low-temperature properties of the large-$U$ Hubbard model
in two and three dimensions from exact series-expansion results for high
temperatures. Convergence
problems and limited available information prevent a direct or Pad\'e-type
extrapolation. We propose a new method of extrapolation, which is
restricted to large $U$ and low hole densities, for which the problem can be
mapped on that of a system of weakly interacting holes. In the new formulation
an extrapolation down to $T=0$ can be obtained, but it can be trusted for the
presently available series data for $\beta t \lesssim 20$ and for hole
densities $n_h \lesssim 0.2$ only. Implications for the magnetic phase
diagram are discussed.
\end{abstract}


\widetext
\section{Introduction}
\label{Intro}
The single-band Hubbard model is presumably the simplest model for
describing the behavior of correlated electrons in a solid. Examples
of its applications are its initial use to describe magnetism in
transition metals, and, most recently, theories of high-temperature
superconductivity. Unfortunately, while it seems likely that for the
latter phenomenon more complex models are needed, even this simple
model is not nearly well understood. For one dimension some rigorous
results are known, but in higher dimensions the main results have been
obtained from Monte Carlo and finite-lattice calculations only.

We are interested in deriving magnetic properties for the case of large $U$ on
a square or simple-cubic lattice. A well known theorem by Nagaoka\cite{nagaoka}
states that a Hubbard model on a bipartite lattice with one hole and at
infinite~$U$ has a ferromagnetic ground state. Many authors have investigated
whether this one point in the phase diagram is part of a whole region of
ferromagnetic behavior. Various methods are being used for this
purpose\cite{summary}, including exact diagonalization of small systems, Monte
Carlo simulations, mean-field, and cluster expansion methods. Two of
us\cite{tenhaaf} as well as various other authors\cite{henderson,expansions}
have used the last method to calculate high-temperature series expansions for
the square and simple-cubic lattices. Expressions have been obtained for
various thermodynamical quantities, such as the free energy, the magnetization,
the magnetic susceptibility, and also for the pair-correlation functions
between the $z$ components of the spin at specific sites. These expressions
show very well convergent behavior for high temperatures ($kT/t \gtrsim 2$).
The aim was to find indications for the onset to ferromagnetic behavior at low
temperatures by extrapolating the results of the series expansions. Indeed,
these indications can be found, as is shown in Refs.\onlinecite{tenhaaf}
and\onlinecite{henderson}. However, predictions for the ground state, based on
these results, are highly unreliable. Due to the fact that we only have five
terms in the series expansions (0th, 2nd, 4th, 6th, and 8th order terms), we
found it impossible to rely on standard extrapolation methods like Pad\'e
approximants. The obtained results for different extrapolations vary too much
to be able to derive any reliable extrapolated value. Henderson et
al.\cite{henderson}\ try to find an indication for the expected divergence in
the uniform
susceptibility
\begin{equation}\label{fmsusc}
 \chi_{\text{fm}}^{\rule[-.5ex]{0cm}{1ex}} =
\beta\left.\frac{\partial^2}{\partial(\beta h)^2}\ln \text{tr}\ e^{-\beta\cal
H}\right|_{h=0}
\end{equation}
by looking for zeros of $\chi_{\text{fm}}^{-1}$. The character of the series
expansion, shown for infinite~$U$ as a function of $\beta t$ in subsequent
approximations in Fig.~\ref{divergence}, is such that
$\chi_{\text{fm}}^{-1}(\beta t)$ is likely to diverge very quickly for $\beta
t>1$, to plus and minus infinity alternatingly. This means that zeros are to be
found for $\beta t\gtrsim 1$ in the fourth- and eighth-order approximations,
but no zeros exist in the second and sixth orders.

We feel that there is no reason to believe that the fourth- and eighth-order
results should be more reliable than the others.

Furthermore, we have also constructed the antiferromagnetic susceptibility
$\chi_{\text{af}}^{\rule[-.5ex]{0cm}{1ex}}$ by including a staggered-field term
in the Hamiltonian, and we have calculated its divergence in the same way as
described above. In Fig.~\ref{contours} we compare the Curie temperature
$T_{\text{C}}$ as a function of the particle density, for various values of
$t/U$, to the N\'eel temperature $T_{\text{N}}$, for calculations up to eighth
order ($T_{\text{C}}$ and $T_{\text{N}}$ are defined by
$\chi_{\text{fm}}^{-1}(n,U,T_{\text{C}})=0$ and
$\chi_{\text{af}}^{-1}(n,U,T_{\text{N}})=0$, respectively). As the N\'eel
temperature is higher than the Curie temperature for the parameters shown, one
should conclude that the system goes into an antiferromagnetic state before the
ferromagnetic transition is reached.
However, regarding the character of the series expansion, as illustrated in
Fig.~\ref{divergence}, it is clear that the plots can not be trusted
qualitatively, let alone quantitatively.

In this paper we will consider a method that does not encounter these problems
of extrapolation to low temperatures. In this method the density of holes is
used as a small parameter. The high-temperature results are expressed in terms
of an effective density of states for holes (as was done before by Brinkman and
Rice\cite{brinkman}), and extended to interactions between hole levels. With
this density of states expressions for the free energy of the thermodynamic
system can be obtained in the whole range of temperatures. In Sec.~\ref{Holes}
we will define a partition function for the holes and express it in terms of an
effective chemical potential for the holes. In Sec.~\ref{Density}
we derive the density of states for non-interacting holes, and we determine its
moments, for infinite $U$. We present an improvement on the non-interacting
hole picture in Sec.~\ref{Interact}, where we consider interacting holes by
introducing a Fermi-liquid-like interaction in energy space. In
Sec.~\ref{Extrap} we show how to use the density of states to calculate zeros
of the inverse susceptibility. Sec.~\ref{FiniteU} deals with the
non-interacting hole approximation applied for finite~$U$. In Sec.~\ref{Phase}
we show our conclusions for the magnetic phase diagram, and we discuss the
method in Sec.~\ref{Discuss}.

\section{Hole formulation}
\label{Holes}

We consider the Hubbard Hamiltonian
\begin{equation}
{\cal H} = {\cal H}_{\text{kin}} + {\cal H}_{\text{local}} -
\mu\sum_{i\sigma} n_{i\sigma}^{\rule[-.5ex]{0cm}{2ex}} -
h\sum_{i\sigma} \sigma n_{i\sigma}^{\rule[-.5ex]{0cm}{2ex}}
\end{equation}
with
\begin{eqnarray}
{\cal H}_{\text{kin}} = -t\sum_{\langle i,j\rangle ,\sigma}
c_{i\sigma}^{\dagger\rule[-.5ex]{0cm}{2ex}}%
c_{j\sigma}^{\rule[-.5ex]{0cm}{2ex}} \\
{\cal H}_{\text{local}} =
U\sum_i n_{i\uparrow}^{\rule[-.5ex]{0cm}{2ex}}%
n_{i\downarrow}^{\rule[-.5ex]{0cm}{2ex}}
\end{eqnarray}
where $t$~is the hopping integral between nearest neighbors, $U$~denotes the
on-site interaction strength, $\mu$~is the chemical potential, and $h$~is the
strength of an external magnetic field. The operator
$c_{i\sigma}^{\dagger\rule[-.5ex]{0cm}{2ex}}$
($c_{i\sigma}^{\rule[-.5ex]{0cm}{2ex}}$) creates (annihilates) a particle with
spin $\sigma$ at site $i$, and $n_{i\sigma}^{\rule[-.5ex]{0cm}{2ex}} =
c_{i\sigma}^{\dagger\rule[-.5ex]{0cm}{2ex}}c_{i\sigma
}^{\rule[-.5ex]{0cm}{2ex}}$ counts
the number of particles with spin $\sigma$ at site $i$.

To investigate the thermodynamic properties we want to calculate the grand
canonical partition function
\begin{equation}
Z_{\text{gr}} = \text{tr}\ e^{-\beta{\cal H}}
\end{equation}
For a system consisting of $N$ sites we can rewrite this as
\begin{equation} \label{gcpfsum}
Z_{\text{gr}} = \sum_{N_s=0}^{2N} e^{\beta\mu N_s}Z_{N_s}
\end{equation}
with $Z_{N_s}$ the canonical partition function for $N_s = N_\uparrow +
N_\downarrow$ particles:
\begin{equation}
Z_{N_s} = \sum_{N_\uparrow=0}^{N_s} e^{\beta h(N_\uparrow-N_\downarrow)}
\sum_j e^{-\beta \varepsilon_j^{(N,N_s,N_\uparrow)}}
\end{equation}
Here $\{\varepsilon_j^{(N,N_s,N_\uparrow)}\}$ is the set of eigenvalues of
${\cal H}_{\text{kin}} + {\cal H}_{\text{local}}$ for $N_\uparrow$~up spins and
$N_\downarrow$~down spins on $N$~sites (note that the~$\varepsilon_j$ are
functions of~$t$ and~$U$ only). Via the grand potential
\begin{equation}\label{Omega}
\Omega = -\frac{1}{\beta} \ln Z_{\text{gr}}
\end{equation}
we can now derive the other thermodynamic quantities by means of the usual
manipulations, e.g. the particle density
\begin{equation} \label{density}
n_s = \frac{\langle N_s\rangle}{N} =
-\frac{1}{N}\frac{\partial\beta\Omega}{\partial\beta\mu}
\end{equation}
or the susceptibility as given by~(\ref{fmsusc}).

In order to approach to lower temperatures in the limit of strong interactions
and near half filling, we are going to express the partition function in terms
of an effective chemical potential for holes. We associate the kinetic part of
the Hamiltonian with the motion of the (dilute) holes, and its magnetic part
with the background of spins. Thus, we have to divide out the spin degrees of
freedom to obtain the canonical partition function for the holes:
\begin{equation} \label{cpfholes}
Z_{N_h}^{\text{h}} \equiv Z_{N-N_h}e^{\beta\epsilon_{\text{hf}}(N-N_h)}
\end{equation}
Here we define the number of holes
\begin{equation}\label{Nh}
 N_h = N - N_s
\end{equation}
and we introduce a parameter $\epsilon_{\text{hf}}$ which can be viewed as the
free energy per spin in the absence of holes (i.e. at half filling; naturally,
$Z_{0}^{\text{h}}\equiv 1$):
\begin{eqnarray} \label{defepsilon_hf}
\epsilon_{\text{hf}} & \equiv & -\frac{1}{N\beta}\ln Z_N  \\ \label{epsilon_hf}
& = & -\frac{1}{\beta}\ln(2\cosh \beta h)
\end{eqnarray}
for infinite~$U$. The grand canonical partition function for the holes then is
\begin{eqnarray} \label{gcpfholes}
Z_{\text{gr}}^{\text{h}} & = &
Z_{\text{gr}}e^{\beta(\epsilon_{\text{hf}}-\mu)N}  \\ \label{gcpfholessum} & =
& \sum_{N_h} Z_{N_h}^{\text{h}}e^{\beta(\epsilon_{\text{hf}}-\mu)N_h}
\end{eqnarray}
suggesting the definition of an effective chemical potential for the holes (Cf.
Eq.~(\ref{gcpfsum})):
\begin{equation}\label{mu_h}
\mu_{\text{h}} \equiv \epsilon_{\text{hf}}-\mu
\end{equation}
With this definition we can rewrite~(\ref{gcpfholes}) as
\begin{equation} \label{basicZZ_h}
 \ln Z_{\text{gr}} = -\beta\mu_{\text{h}}N + \ln Z_{\text{gr}}^{\text{h}}
\end{equation}

Note that expression~(\ref{epsilon_hf}) for $\epsilon_{\text{hf}}$ is exactly
true only in the case of infinite~$U$, as the interaction then prevents
particles from occupying the same site. Note also that we do not define the
number of holes as the number of sites where no particles are present (a
definition which seems obvious), because the interpretation of
Eqs.~(\ref{cpfholes}) and~(\ref{gcpfholessum}) would then become problematic
for finite~$U$. However, if $U$ is very large, as we assume, a pair of
electrons  located on the same site causes a very high energy, and the
contribution of the corresponding hole to the kinetic part of the Hamiltonian
is some orders of magnitude smaller than the contribution of a `real'
(non-removable) hole. Therefore we will use~(\ref{Nh}) also in the case of
large, finite~$U$, and we will show that this leads to terms to be added to the
expressions for infinite~$U$ of order $\case{1}{U}$ or higher.

\section{Construction of the density of states for infinite~U}
\label{Density}
Let us consider a system near half filling, with, for simplicity, infinitely
strong coupling $U$ (the case of finite~$U$ will be treated in
Sec.~\ref{FiniteU}). We assume that the system can be described in terms of the
kinetic energy of non-interacting dilute holes and the magnetic energy of the
background particles. We define
\begin{equation}\label{defrho}
 \frac{Z_1^{\text{h}}}{N} \equiv \int d\varepsilon\rho(\varepsilon,\beta
h)e^{-\beta t\varepsilon}
\end{equation}
where $\rho(\varepsilon,\beta h)$ is the spectral distribution of the energy
levels of one hole in an otherwise half-filled system. We take $\rho$ to be
normalized to one.

Although we said before that we divide out the magnetic degrees of freedom
in the spin background, there is still a dependence of $\rho$ on the magnetic
field $h$. It is not easy to see how the hole motion depends on the field
exactly, but one can easily understand why this dependence exists: a magnetic
field influences the distribution of the spin background, which in turn
determines the behavior of the hole. The hole motion depends on the field only
indirectly, and the mechanism that governs the hole dynamics can in fact be
much better described in terms of the average magnetization of the spin
background than in terms of the field. It is important to understand that, in
this picture, one has to treat the spin background as if it were at half
filling, with the dilute holes subjected to its magnetization. Therefore we
change variables at this level from $\beta h$ to the magnetization per spin
$m$. This change is easily performed by a Legendre transformation
\begin{equation}\label{legendre}
 \bar{\epsilon}_{\text{hf}}(m) = \epsilon_{\text{hf}}(\beta h) + mh
\end{equation}
yielding
\begin{equation}\label{rhom}
 \frac{Z_1^{\text{h}}}{N} \equiv \int
d\varepsilon\bar{\rho}(\varepsilon,m)e^{-\beta t\varepsilon}
\end{equation}
where $\bar{\rho}(\varepsilon,m)$ is obtained from $\rho(\varepsilon,\beta h)$
via
\begin{equation}\label{mh}
 m = -\frac{\partial \epsilon_{\text{hf}}}{\partial h}
\end{equation}
which becomes $m=\tanh (\beta h)$ for infinite~$U$.

With this definition we can write down a first approximation for the grand
canonical partition function. A one-hole level can be occupied, with a
Boltzmann weight $e^{-\beta t\varepsilon}$ (where we write $t\varepsilon$ to
make the dependence of the hole energy on $t$ explicit), or it can be
unoccupied, in which case there is an electron in the system with Boltzmann
weight $e^{-\beta\mu_{\text{h}}}$ (with the magnetic energy included in
$\mu_{\text{h}}$). Thus, in the case of non-interacting holes we have (dropping
the $m$-dependence of $\bar{\rho}$)
\begin{equation}\label{Zgrapprox}
 \ln Z_{\text{gr}} = N\int d\varepsilon\bar{\rho}(\varepsilon)\ln(e^{-\beta
t\varepsilon} + e^{-\beta\mu_{\text{h}}})
\end{equation}
or equivalently, using~(\ref{basicZZ_h}),
\begin{equation}\label{Zgrholrho}
 \ln Z_{\text{gr}}^{\text{h}} = N\int
d\varepsilon\bar{\rho
}(\varepsilon)\ln\left(1+e^{-\beta(t\varepsilon-\mu_{\text{h}})}\right)
\end{equation}

This equation becomes exact in a one-dimensional system, as in that case the
holes can not disturb the magnetic background of the particles, thus being
really non-interacting, and also in a ferromagnetic system (at $m=\pm 1$), for
similar reasons. In other, higher-dimensional systems (\ref{Zgrholrho}) is only
correct to first order in $e^{\beta\mu_{\text{h}}}$. We make an expansion of
the right-hand side with respect to the small parameter
$e^{\beta\mu_{\text{h}}}$ to obtain
\begin{equation} \label{Zgrholexp}
 \ln Z_{\text{gr}}^{\text{h}} = N\int
d\varepsilon\bar{\rho}(\varepsilon)\left(e^{\beta\mu_{\text{h}}}e^{-\beta
t\varepsilon} - \frac{1}{2}e^{2\beta\mu_{\text{h}}}e^{-2\beta t\varepsilon} +
\ldots\right)
\end{equation}
Comparing this to a similar expansion of the logarithm of
Eq.~(\ref{gcpfholessum}) we see that this is consistent with the definition of
the density of states in the first-order term.

Now, as an illustration of the calculation, let us have a look at the form that
Eq.~(\ref{Omega}) actually takes when evaluating it for this system by means of
a cluster expansion method. In Ref.\onlinecite{tenhaaf} a general formula is
presented for finite~$U$, which for infinite~$U$ reduces to
\begin{equation}\label{ZgrtH}
 \frac{\ln Z_{\text{gr}}}{N} = \ln Z_{\text{gr},0} + \sum_{n=1}^\infty(\beta
t)^{2n}\sum_{m=0}^{2n-1}%
\sum_{\,\,l=-m}^{m}\frac{\Omega_{m,l}^{(n)}e^{\beta(m\mu+lh)}
}{(Z_{\text{gr},0})^{2n}}
\end{equation}
Here $Z_{\text{gr},0}$ denotes the partition function for a system consisting
of only one site:
\begin{equation}
 Z_{\text{gr},0} = 1 + 2e^{\beta\mu}\cosh(\beta h)
\end{equation}
The $\Omega_{m,l}^{(n)}$ are coefficients, which can e.g. be calculated by
means of a cluster expansion method. By substituting $\mu_{\text{h}}$ for
$\mu$, using~(\ref{mu_h}) and~(\ref{epsilon_hf}), and expanding in the small
parameter $e^{\beta\mu_{\text{h}}}$, we can obtain an expression for the grand
potential for the holes again, now in the form of a series expansion:
\begin{equation}\label{Zgrexp}
 \frac{\ln Z_{\text{gr}}^{\text{h}}}{N} = \sum_{N_h=1}^\infty
e^{N_h\beta\mu_{\text{h}}}\sum_{n=0}^\infty (\beta t)^{2n} \Omega(N_h,n,h)
\end{equation}
where
\begin{equation}
 \Omega(p,0,h) = \frac{(-1)^{(p-1)}}{p}
\end{equation}
and
\begin{equation}
 \Omega(p,n,h) =
\sum_{m=0}^{2n-1}\sum_{\,\,l=-m}^m\!\left(\!\begin{array}{c
}p+m-1\\p+m-2n\end{array%
}\!\right)\frac{(-1)^{p}\Omega_{m,l}^{(n)}e^{\beta lh}}{\left(-2\cosh
(\beta h)\right)^m}
\end{equation}
for $n\not=0$.
Finally, we obtain a relation between the coefficients $\Omega(1,n,h)$ and the
moments of $\rho(\varepsilon)$ ($=\rho(\varepsilon,\beta h)$) by
expanding~(\ref{defrho}) in powers of $\beta t$:
\begin{equation}\label{Z1holemoments}
\frac{Z_1^{\text{h}}}{N} = \sum_{n=0}^\infty (\beta t)^n \frac{(-1)^n}{n!}\int
d\varepsilon\rho(\varepsilon)\varepsilon^n
\end{equation}
Thus, we see from Eq.~(\ref{Z1holemoments}) and the first-order term
in~(\ref{Zgrexp}) that we have
\begin{equation}
 \int d\varepsilon\rho(\varepsilon)\varepsilon^{2n} = (2n)!\Omega(1,n,h)
\end{equation}
for the even moments of $\rho$, all odd moments being zero.
Although we have restricted ourselves to the case of infinite~$U$ here, this
expression can easily be extended for the case of finite~$U$, as we will see in
Sec.~\ref{FiniteU}. $U$ then enters the equation as a parameter at the
right-hand side.

For infinite~$U$ there is another, faster way to calculate these moments. They
can then be expressed directly in the number of possible paths in state space
for a system with one hole. This has been done by Brinkman and
Rice\cite{brinkman}, who calculated the first 10 moments of the density of
states for ferromagnetic, antiferromagnetic and paramagnetic spin backgrounds
on a simple cubic lattice, and by Meshkov and Berkov\cite{meshkov}, who
calculated 16 moments for the same spin backgrounds on a square lattice. In
Appendix~A, we outline a method which enables us to enumerate the paths in an
efficient way, and by which we have extended the results for the square lattice
to 22 moments. These moments are presented in Table~\ref{moments} for $m=0$ and
$m=1$, corresponding to a paramagnetic and ferromagnetic system, respectively.
We have also extended the results for the simple cubic lattice, to 16 moments;
they are available on request.

We now approximate $\bar{\rho}(\varepsilon)$ by a polynomial which we fit with
the moments. In this way we calculate an approximation for the density of
states, to different orders, in order to get an impression of the convergence
of subsequent approximations. In Figs.~\ref{density_of_states}a
and~\ref{density_of_states}b we show the result for a paramagnetic and a
ferromagnetic system.

For $m=1$ the exact density of states is known:
\begin{equation}
 \bar{\rho}(\varepsilon) =
\frac{1}{2\pi^2}\,K\!\left(1-\left(\frac{\varepsilon}{4}\right)^2\right)
\end{equation}
with $K$ the complete elliptic integral of the first kind. It has an integrable
singularity at $\varepsilon=0$. This is difficult to approximate and causes
some  oscillations away from $\varepsilon=0$. Convergence towards the exact
result is rather good. For $m=0$ convergence is very good, as from 14th order
on the difference between subsequent approximations becomes very small.

Meshkov and Berkov fit the density of states by postulating that the integral
of $\bar{\rho}^2$ be minimal (`smoothness' criterion), using a discretized
$\bar{\rho}$. They claim that this method gives faster convergence then a
polynomial fit. Comparing their results for the ferromagnetic density of states
with the exact result and the results presented here, however, one may question
that claim. We feel that the polynomial fits, when using an equal number of
moments, give similar or even better results, which are also easier to handle
in further calculations.

Before calculating various quantities which can tell us something about the
low-temperature properties of the system, we will in the next section consider
a method to improve the approximation of the density of states by including
interactions between the holes.

\section{Interacting holes}
\label{Interact}
The crucial question is to see for which domain of hole densities the
assumption of independent holes is justified. This range can be determined from
an estimate of the interactions between the holes. Very similar to the theory
of the classical dilute gas\cite{clasgas}, the interaction can be deduced from
the two-hole partition function as defined by~(\ref{cpfholes}) for $N_h=2$. It
is a matter of choice how to represent the hole interaction. One could think of
a spatial representation, but one must realize that in this strongly quantal
system the interaction is non-local, which complicates the transparancy of the
representation substantially. Having the one-hole system represented by a
density of states it is natural here to choose an interaction between energy
levels.
First we formulate the interaction in terms of discrete levels and then we take
the continuum limit as in Eq.~(\ref{Zgrapprox}). The discrete version of this
expression can be written in terms of levels $\varepsilon_i$, distributed
according to the density $\bar{\rho}(\varepsilon_i)$:
\begin{equation}
 Z_{\text{gr}}^{(1)} =
e^{-\beta\mu_hN}\sum_{\{n_i\}}e^{-\beta\sum_i(t\varepsilon_i-\mu_h)n_i}
\end{equation}
where $\{n_i\}$ with $n_i=0,1$ is the occupation of the levels $\varepsilon_i$.
We have given the expression a superindex 1 to indicate that
$Z_{\text{gr}}^{(1)}$ matches $Z_{\text{gr}}$ up to the one-hole terms. The
next approximation can be of the form
\begin{equation}\label{gcpf2def}
 Z_{\text{gr}}^{(2)} =
e^{-\beta\mu_hN}\sum_{\{n_i\}}e^{-\beta\left(\sum_i(t\varepsilon_i-
\mu_h)n_i+\sum_{(i,j)}f_{ij}n_in_j\right)}
\end{equation}
where $f_{ij}$ accounts for the interaction between the levels $\varepsilon_i$
and $\varepsilon_j$. The second term in the exponent is a sum over all pairs of
levels $(i,j)$. In the energy space a distance between levels does not seem to
be a measure for the strength of the interaction as in real space, where
interactions usually decay sufficiently fast with the distance, such that the
sum over pairs does not increase with the square of the number of elements, but
only linearly as is necessary for a thermodynamic system. In order to make the
exponent in~(\ref{gcpf2def}) of the correct thermodynamic behavior the
interaction should therefore decrease with the size of the system as
\begin{equation}\label{phidef}
 f_{ij}=\frac{t}{N}\phi_{ij}
\end{equation}
with $\phi_{ij}$ of order unity. An additional advantage of~(\ref{phidef}) is
the fact that interactions of this type can be handled rigorously in the
thermodynamic limit by the mean-field theory\cite{meanf}. Thus we can write
\begin{equation}\label{gcpf2}
 \ln Z_{\text{gr}}^{(2)} = -\beta\mu_hN +
\sum_i\ln(1+e^{-\beta\tilde{\varepsilon_i}}) + \frac{\beta
t}{N}\sum_{(i,j)}\phi_{ij}n(\tilde{\varepsilon_i})n(\tilde{\varepsilon_j})
\end{equation}
\noindent where the $\tilde{\varepsilon_i}$ are the shifted energy levels
\begin{equation}\label{epsshift}
 \tilde{\varepsilon_i} =
t\varepsilon_i-\mu_h+\frac{t}{N}\sum_{j\not=i}\phi_{ij}n(\tilde{\varepsilon_j})
\end{equation}
and $n(\tilde{\varepsilon})$ is the fermi occupation number
\begin{equation}\label{fermifac}
 n(\tilde{\varepsilon})=\frac{1}{1+e^{\beta\tilde{\varepsilon}}}
\end{equation}
Now the interaction $\phi_{ij}$ must be chosen such that $Z_{\text{gr}}^{(2)}$
produces the correct two-hole partition function. Expanding
Eq.~(\ref{gcpf2def}) with respect to the number of holes
\begin{equation}
 N_h = \sum_in_i
\end{equation}
and using Eq.~(\ref{gcpfholes}) we find
\begin{equation}\label{cpf2holes}
 Z_2^{\text{h}} = \sum_{(i,j)}e^{-\beta
t(\varepsilon_i+t\varepsilon_j+\frac{1}{N}\phi_{ij})}
\end{equation}
In our high-temperature expansion we have no direct information on
$Z_2^{\text{h}}$, but we have the coefficient of the second-order term in the
hole expansion of $\ln Z_{\text{gr}}^{\text{h}}$ (Cf.~(\ref{gcpfholessum})),
which is
\begin{equation}\label{U2def}
 U_2 = Z_2^{\text{h}}-\frac{1}{2}(Z_1^{\text{h}})^2
\end{equation}
Note that this expression is of order~$N$, and not of order $N^2$ as are both
terms on its right-hand side. Using~(\ref{cpf2holes}) and the corresponding
expression for $Z_1^{\text{h}}$ we may equate
\begin{equation}\label{U2exp}
 U_2 = \sum_{(i,j)}e^{-\beta t(\varepsilon_i+\varepsilon_j +
\frac{1}{N}\phi_{ij})} - \frac{1}{2}\sum_{i,j}e^{-\beta
t(\varepsilon_i+\varepsilon_j)}
\end{equation}
For $N\rightarrow\infty$ we may write
\begin{equation}\label{U2phi}
 U_2 = -\frac{\beta t}{N}\sum_{(i,j)}e^{-\beta
t(\varepsilon_i+\varepsilon_j)}\phi_{ij} - \frac{1}{2}\sum_ie^{-2\beta
t\varepsilon_i}
\end{equation}
and we see that $U_2$ is indeed of order $N$ by virtue of~(\ref{phidef}). Note
that the terms $i=j$ in the second term of~(\ref{U2exp}) are not compensated by
the first term. The second term in~(\ref{U2phi}) gives the ideal-gas term of
the hole system on the two-hole level.

Since we have moments of $U_2$ by our high-temperature expansions, and also the
last term in~(\ref{U2phi}) is known from our one-hole density of states, it is
convenient to split $U_2$ into an interacting and an ideal part
\begin{equation}
 U_2 = U_2^{\text{int}} + U_2^{\text{id}}
\end{equation}
with
\begin{eqnarray}
 U_2^{\text{id}} & = & -\frac{1}{2}\sum_ie^{-2\beta t\varepsilon_i} \nonumber\\
 & = & -\frac{N}{2}\int d\varepsilon\bar{\rho}(\varepsilon)e^{-2\beta
t\varepsilon}
\end{eqnarray}
(Cf.~(\ref{Zgrholexp})). $\phi_{ij}$ must then be determined from
$U_2^{\text{int}}$. In a continuous version the equation for
$\phi(\varepsilon,\varepsilon^\prime)$ becomes
\begin{equation}\label{U2int}
 U_2^{\text{int}} = -\frac{\beta t}{2}N\int
d\varepsilon\bar{\rho}(\varepsilon)\int
d\varepsilon^\prime\bar{\rho}(\varepsilon^\prime)e^{-\beta
t(\varepsilon+\varepsilon^\prime)}\phi(\varepsilon,\varepsilon^\prime)
\end{equation}
This relation is not strong enough to yield a unique
$\phi(\varepsilon,\varepsilon^\prime)$, in the same way as the second virial
coefficient of a classical gas is not sufficient to determine the interaction
potential. The freedom in choice will be reflected upon the efficiency of the
program to determine the higher-order interactions. We have chosen to have the
dependence of $\phi(\varepsilon,\varepsilon^\prime)$ only on the sum variable
$\varepsilon+\varepsilon^\prime$, and we approximate it by a polynomial:
\begin{equation}\label{phi_ldef}
 \phi(\varepsilon,\varepsilon^\prime) =
\sum_l\phi_l(\varepsilon+\varepsilon^\prime)^l
\end{equation}
Equating moments in~(\ref{U2int}) and in
\begin{equation}
 U_2^{\text{int}} = \sum_k(U_2^{\text{int}})_k(\beta t)^k
\end{equation}
we have
\begin{equation}\label{U2int_k}
 \left(U_2^{\text{int}}\right)_k =
\sum_l\!\frac{(-1)^kN}
{2(k-1)!}\!\int\!\!d\varepsilon\bar{\rho%
}(\varepsilon)\!\int\!\!d\varepsilon^\prime\bar{\rho%
}(\varepsilon^\prime)(\varepsilon+\varepsilon^\prime)^{k-1+l}\phi_l
\end{equation}
Because we are working on a bipartite lattice, all odd moments of
$U_2^{\text{int}}$ are identically zero. Hence $k$ is even, and as also
$\bar{\rho}(\varepsilon)$ has only even moments, the combination $k-1+l$ must
be even and therefore the sums in Eqs.~(\ref{phi_ldef}) and~(\ref{U2int_k})
contain only odd $l$. The set of equations~(\ref{U2int_k}) for a finite number
of the moments $\left(U_2^{\text{int}}\right)_k$ determines an equal number of
coefficients $\phi_l$. We have computed $\left(U_2^{\text{int}}\right)_k$ for
the square lattice at $U=\infty$, up to $k=12$. This involves 6 terms ($k=2, 4,
\ldots, 12$) and so we can determine 6 values $\phi_1, \phi_3, \ldots,
\phi_{11}$. In the equations we thus need $k-1+l=12-1+11=22$ as the highest
moment of $\bar{\rho}(\varepsilon)$, which is just the number of moments we
have determined. The values of $\frac{1}{N}\left(U_2^{\text{int}}\right)_k$ are
given in Table~\ref{intcoef}.
For the ferromagnetic system ($m=1$) these coefficients are zero, as the holes
do not interact in that case.

Finally we give the continuum form of the expressions~(\ref{gcpf2})
and~(\ref{epsshift}) for the grand potential:
\begin{equation} \label{Zgrint}
 \ln Z_{\text{gr}}^{(2)} = -\beta\mu_hN + N\int
d\varepsilon\bar{\rho}(\varepsilon)\ln(1+e^{-\beta\tilde{\varepsilon}}) +
\frac{\beta t}{2}N\int d\varepsilon\int
d\varepsilon^\prime\bar{\rho}(\varepsilon)n(\tilde{\varepsilon})\bar{\rho%
}(\varepsilon^\prime)n(\tilde{\varepsilon%
}^\prime)\phi(\varepsilon,\varepsilon^\prime)
\end{equation}
\noindent with
\begin{equation} \label{epstilde}
 \tilde{\varepsilon} = t\varepsilon-\mu_h+t\int
d\varepsilon^\prime\bar{\rho%
}(\varepsilon^\prime)\phi(\varepsilon,\varepsilon^\prime)n(\tilde{\varepsilon%
}^\prime)
\end{equation}

\section{Inverse susceptibility}
\label{Extrap}
We return to the uniform susceptibility
\begin{equation}
 \chi_{\text{fm}}^{\rule[-.5ex]{0cm}{1ex}} = \left.\frac{\partial M}{\partial
h}\right|_{h=0} 
\end{equation}
with $M$ the total magnetization of the system. As before, we try to find
indications of divergences of $\chi_{\text{fm}}^{\rule[-.5ex]{0cm}{1ex}}$,
which should be related to second-order phase transitions between a
paramagnetic and a ferromagnetic state. It is usually more convenient to
express this by stating that the inverse susceptibility must be zero
\begin{equation} \label{stoner}
 \chi_{\text{fm}}
^{-1\rule[-.5ex]{0cm}{1ex}} = 0
\end{equation}
and to study
\begin{equation}
  \chi_{\text{fm}}^{-1\rule[-.5ex]{0cm}{1ex}} = \left.\frac{\partial
h}{\partial M}\right|_{M=0}
\end{equation}
or
\begin{equation}\label{invsus}
 \beta N\chi_{\text{fm}}^{-1\rule[-.5ex]{0cm}{1ex}} = \left.\frac{\partial
\beta h}{n_{\text{s}}\partial m}\right|_{m=0}
\end{equation}
where $m$ is the magnetization per spin as defined in Sec.~\ref{Density}.
In order to find an expression for $h$, to be able to calculate (\ref{invsus}),
we construct a generalized (Landau like) free energy
\begin{equation} \label{deffreeen}
 \varphi(n_{\text{s}},\mu,m,h) = -\frac{1}{\beta N}\ln Z_{\text{gr}} + \mu
n_{\text{s}} - hmn_{\text{s}}
\end{equation}
where $\ln Z_{\text{gr}}$ is given by~(\ref{Zgrapprox}).
$\varphi$ has to be minimized with respect to $\mu$ and $m$ at fixed particle
density $n_{\text{s}}$ and field $h$, to obtain the free energy. Note that this
$h$ is {\em not\/} the same field as we used before in Sec.~\ref{Density}.
There we interpreted $h$ as a field that is felt only by the spins in the
background, whereas now we obtain the physical external field that would be
necessary to yield the given magnetization. Of course, in the case of a finite
number of holes (the limit of half filling), these fields are the same, as we
will see in the resulting expressions. Note also that, due to the definition of
$m$ as the magnetization per spin, its conjugated variable is $hn_{\text{s}}$,
not $h$.

We can rewrite (\ref{deffreeen}) using~(\ref{mu_h}) and~(\ref{legendre}):
\begin{equation}
 \beta\varphi = \beta\bar{\epsilon}_{\text{hf}}n_{\text{s}} +
\beta\mu_{\text{h}}n_{\text{h}} - \frac{1}{N}\ln Z_{\text{gr}}^{\text{h}}
\end{equation}
where we can interpret the first term as the contribution of the background of
spins, and the other terms as the contribution of the holes.
Minimization leads to the following equations:
\begin{eqnarray} \label{holdens}
 n_{\text{h}} & = & \int
d\varepsilon\bar{\rho}(\varepsilon)\frac{1}
{1+e^{\beta(t\varepsilon-\mu_{\text{h}})}} \\
 \label{field}
 \beta h & = & \beta h_{\text{hf}} - \int
d\varepsilon\frac{\partial\bar{\rho}(\varepsilon)}{n_{\text{s}}\partial
m}\ln\left(1+e^{-\beta(t\varepsilon-\mu_{\text{h}})}\right)
\end{eqnarray}
with
\begin{equation} \label{hffield}
 h_{\text{hf}} = \frac{\partial\bar{\epsilon}_{\text{hf}}}{\partial m}
\end{equation}
The expression for the inverse uniform susceptibility (\ref{invsus}) then
becomes
\begin{equation} \label{invsusm}
 \beta N\chi_{\text{fm}}^{-1\rule[-.5ex]{0cm}{1ex}} = \left.\frac{\partial\beta
h_{\text{hf}}}{n_{\text{s}}\partial m}\right|_{m=0} - \int
d\varepsilon\left.\frac{\partial^2\bar{\rho%
}(\varepsilon)}{n_{\text{s}}^2\partial
m^2}\right|_{m=0}\ln\left(1+e^{-\beta(t\varepsilon-\mu_{\text{h}})}\right)
\end{equation}
This can be rewritten in terms of $\rho(\varepsilon)$, using the legendre
transform~(\ref{legendre}) (thus $\bar{\rho}(\varepsilon,m) =
\rho(\varepsilon,\beta h_{\text{hf}})$):
\begin{equation} \label{invsush}
 \beta N\chi_{\text{fm}}^{-1\rule[-.5ex]{0cm}{1ex}} = \left.\frac{\partial\beta
h_{\text{hf}}}{n_{\text{s}}\partial m}\right|_{m=0}\left(1 -
\left.\frac{\partial\beta h_{\text{hf}}}{n_{\text{s}}\partial
m}\right|_{m=0}\int d\varepsilon\left.\frac{\partial^2\rho(\varepsilon,\beta
h_{\text{hf}})}{\partial(\beta
h_{\text{hf}})^2}\right|_{h_{\text{hf}}=0}%
\ln\left(1+e^{-\beta(t\varepsilon-\mu_{\text{h}})}\right)\right)
 \end{equation}
Note that $m=0$ is equivalent to $h_{\text{hf}}=0$, and that, for reasons of
symmetry, the first derivative of $\rho$ with respect to $h_{\text{hf}}$
vanishes at $h_{\text{hf}}=0$.

According to~(\ref{stoner}) we want to find values of $n_{\text{h}}$ and $\beta
t$ for which the right-hand side of~(\ref{invsush}) is zero, with
$n_{\text{h}}$ fixed by Eq.~(\ref{holdens}). One can easily verify that, for
infinite~$U$, we have $\beta h_{\text{hf}}(m) = \arctan(m)$, so
putting~(\ref{invsush}) to zero gives
\begin{equation} \label{tobesolved}
 \int d\varepsilon\left.\frac{\partial^2\rho(\varepsilon,\beta
h_{\text{hf}})}{\partial(\beta
h_{\text{hf}})^2}\right|_{h_{\text{hf}}=0}%
\ln\left(1+e^{-\beta(t\varepsilon-\mu_{\text{h}})}\right) = 1 - n_{\text{h}}
\end{equation}
This equation can be solved by an iterative procedure to calculate the value of
$\mu_{\text{h}}$ for a given value of $\beta t$. The density of states
$\rho(\varepsilon)$, necessary to calculate $n_{\text{h}}$ according
to~(\ref{holdens}), is determined from its moments as described in
Sec.~\ref{Density}, and its second derivative is calculated in a similar way.

To include the interaction described in Sec.~\ref{Interact}, one should use the
grand potential as given in~(\ref{Zgrint}) rather than the non-interacting hole
approximation of~(\ref{Zgrapprox}). The final equation, equivalent
to~(\ref{invsush}), then involves one extra term which contains the second
derivative with respect to $\beta h_{\text{hf}}$ of the interaction $\phi$. We
give a derivation of this equation in Appendix~B.

In Fig.~\ref{curie2infU} we show Curie temperatures for the square-lattice
Hubbard model at infinite~$U$, in three different approximations: (a) The
non-interacting hole approximation, with $\rho$ determined by interpolation
from 8 of its moments (of which 4 moments are non-zero); (b) The same but with
$\rho$ determined from 22 (11 non-zero) moments; and (c) The interacting-hole
approximation, with $\rho$ determined from 22 moments and $\phi$ from 12 (5
non-zero) interaction coefficients.

One can see that the difference between the 8th- and the 22nd-order
non-interacting approximations is small. In both approximations, ferromagnetism
is stable against paramagnetism for $n_{\text{h}} \lesssim 0.27$, at low $T$.
The interaction does not change this picture very much. It slightly enhances
the stability of the ferromagnetic state, up to $n_{\text{h}} \lesssim 0.29$.
The difference between the non-interacting and the interacting approximations
becomes larger with increasing hole density, as expected. Numerically, the
results agree very well for $n_{\text{h}}\lesssim 0.06$.

In the next section we will treat the case of finite~$U$. We have been able to
calculate 8 moments of the density of states in that case, thus we can do an
eighth-order approximation at the most. One can then calculate merely two
coefficients $\phi_l$ of the interaction, resulting in an approximation of the
interaction which is rather crude. We have seen that the picture in the
non-interacting hole approximation is qualitatively the same as the one in the
interacting-hole approximation, in eighth order already. For small
$n_{\text{h}}$ it agrees rather well also numerically. Therefore, we will not
include the interaction in the following calculations.

\section{The non-interacting hole approximation for finite~U}
\label{FiniteU}
As we pointed out before, at finite~$U$, excitations in the spin background
become possible due to the creation of pairs of electrons with opposite spin at
the same site. This means that extra empty sites are created, and thus the
number of empty sites in the system is no longer fixed. Taking $U$ large,
however, we can consider the contributions to the partition function due to
these excitations to be small corrections of the infinite-$U$ system, and we
can neglect the terms that would arise from permanently present electron pairs.
To do this, we consider the grand potential of the Hubbard model on a square
lattice up to the second-order term (taken from Ref.\onlinecite{tenhaaf}; note
that $h$ is the parameter in the Hamiltonian here, not the physical magnetic
field we discussed in the previous section):
\begin{eqnarray} \label{Om2sum}
 -\frac{\beta\Omega}{N} & = & \ln\left(1+2e^{\beta\mu}\cosh(\beta h) +
e^{2\beta\mu - \beta U}\right) + \nonumber \\ & + & (\beta
t)^2\frac{4e^{\beta\mu}(1+e^{2\beta\mu-\beta U})\cosh(\beta h) + \frac{8}{\beta
U}e^{2\beta\mu}(1 - e^{-\beta U})}{\left(1+2e^{\beta\mu}\cosh(\beta h) +
e^{2\beta\mu - \beta U}\right)^2} + \ldots
\end{eqnarray}
In this expression, we will neglect the terms that contain the exponential of
$-\beta U$, but we keep terms that are proportional to a power of $1/U$. This
precisely distinguishes the terms that are due to permanent electron pairs,
which cause an energy $\beta U$, from those due to temporary excitations in a
system where otherwise no double occupancies are present. It is necessary to
make this approximation, as the exponential terms can not be treated in this
method. However, it can be seen easily that these terms are always
exponentially smaller than other terms in the expansion, and thus that this
approximation is justified.

First we consider the case of half filling, where we have $\mu = U/2$:
\begin{equation}
 -\frac{\beta\Omega_{\text{hf}}}{N} = \ln\left(2+2e^{\beta U/2}\cosh(\beta
h)\right) + (\beta t)^2\frac{8e^{\beta U/2}\cosh(\beta h) + \frac{8}{\beta
U}(e^{\beta U} - 1)}{\left(2+2e^{\beta U/2}\cosh(\beta h)\right)^2} + \ldots
\end{equation}
Here we can neglect all but the most important terms at large~$U$, i.e. we only
take the terms containing the highest power of $e^{\beta U}$, to get
\begin{equation}
 -\frac{\beta\Omega_{\text{hf}}}{N} = \frac{\beta U}{2} + \ln\left(2\cosh(\beta
h)\right) + (\beta t)^2\frac{2}{(\beta U)\left(\cosh(\beta h)\right)^2} +
\ldots
\end{equation}
By definition, this expression must be equal to $\frac{\beta U}{2} +
\frac{1}{N}\ln Z_{\text{N}}$, so using the definition~(\ref{defepsilon_hf}) for
$\epsilon_{\text{hf}}$ we get
\begin{equation} \label{epsilon_hfU}
 -\beta\epsilon_{\text{hf}} = \ln\left(2\cosh(\beta h)\right) + (\beta
t)^2\frac{2}{(\beta U)\left(\cosh(\beta h)\right)^2} + \ldots
\end{equation}
and we see that this is indeed a correction of order $\frac{1}{U}$ on
Eq.~(\ref{epsilon_hf}). Note that we obtain the same result if we first omit
the $e^{-\beta U}$ terms in~(\ref{Om2sum}), and only then substitute $U/2$ for
$\mu$. This once more supports our statement that these terms may be neglected.

Off half filling, we have to rewrite~(\ref{Om2sum}) (without the $e^{-\beta U}$
terms) in terms of the effective chemical potential $\mu_{\text{h}}$ for the
holes, as defined by Eq.~(\ref{mu_h}), but now containing the corrected
$\epsilon_{\text{hf}}$ as given by~(\ref{epsilon_hfU}). For simplicity, we do
this in a few steps. First, we substitute the chemical potential for the holes
without the correction terms, as in Sec.~\ref{Density}. Then we expand the
logarithm and the numerators with respect to the exponential of this chemical
potential. Finally, we include the corrected $\mu_{\text{h}}$ by expanding the
exponentials with respect to the correction terms. Thus, we obtain for the
grand potential
\begin{equation}
 -\frac{\beta\Omega}{N} = -\beta\mu_{\text{h}} +
e^{\beta\mu_{\text{h}}}\left(1+(\beta t)^2\left[2-\frac{2}{(\beta
U)\left(\cosh(\beta h)\right)^2}\right]+\ldots\right)+\ldots
\end{equation}

The coefficient of $e^{\beta\mu_{\text{h}}}$ in this expression again
determines the moments of the distribution $\rho(\varepsilon,\beta h)$, as
described in Sec.~\ref{Density}. Of course these are now functions of $\beta
U$. In Table~\ref{Umoments} we give the moments that we have been able to
derive from  the series expansion data, for the square lattice, at $h=0$. Note
that the moments for $h=\infty$ are the same as in the case of infinite~$U$
(Table~\ref{moments}), because $U$ has no significance in a system where all
spins point in the same direction.

We can now apply the method described in Sec.~\ref{Extrap} to calculate Curie
temperatures for finite~$U$. One has to realize, though, that at half filling
the inverse susceptibility depends on the temperature, which was not the case
for infinite~$U$. Due to the excitations we get corrections of the type
$\beta t^2/U$, thus we still have a series expansion in the parameter~$\beta
t$. The coefficients in this expansion are suppressed by large factors $\beta
U$, however, and the range of convergence of the expansion is $\beta t\lesssim
30$ or further, depending on the value of $\beta U$. Thus, we may hope that
convergence is good enough in the region where we expect to find solutions
of~(\ref{stoner}). We give the full expression for the inverse susceptibility
at half filling, for the square lattice and up to the $(\beta t)^8$ terms:
\begin{eqnarray} \label{invsushf}
 \beta N\chi_{\text{hf}}^{-1} = \left.\frac{\partial\beta
h_{\text{hf}}}{\partial m}\right|_{m=0} = 1 & + & \frac{4(\beta t)^2}{(\beta
U)} + \frac{8\left(-2+\beta U\right)(\beta t)^4}{(\beta U)^3} + \nonumber \\ &
+ & \frac{\left(1131-648\beta U+32(\beta U)^2\right)(\beta t)^6}{3(\beta U)^5}
+ \nonumber \\
 & + & \frac{\left(-9129+6296\beta U-1132(\beta U)^2+4(\beta U)^3\right)(\beta
t)^8}{(\beta U)^7}
\end{eqnarray}
We have checked that~(\ref{invsushf}) does not become zero for any value of
$\beta t$ and $\beta U$. Therefore we expect no transition from a paramagnetic
to a ferromagnetic state in the half-filled system. Thus, we only have to
consider the second factor on the right-hand side of~(\ref{invsush}), which
vanishes at
\begin{equation}\label{tobesolvedU}
 \int d\varepsilon\left.\frac{\partial^2\rho(\varepsilon,\beta
h)}{\partial(\beta
h)^2}\right|_{h=0}\!\!\ln\left(1+e^{-\beta(t\varepsilon-\mu_{\text{h}})}\right)
= (1 - n_{\text{h}})\frac{\chi_{\text{hf}}}{\beta N}
\end{equation}
We show the results for the square and the simple cubic lattices in the next
section.

\section{Magnetic-phase diagram}
\label{Phase}
We have used the theory described above to calculate Curie temperatures for the
square and simple cubic lattices. For both lattices, we find a surface of Curie
temperatures in the $n_{\text{h}}$--$\frac{t}{U}$--$T$ diagram. In
Figs.~\ref{square} and~\ref{sc} we display these results in various ways.

In Figs.~\ref{square}a and~\ref{sc}a, contours of fixed Curie temperature are
plotted in the $n_{\text{h}}$--$\frac{t}{U}$ plane. In the range of
temperatures up to about $\frac{kT}{t}=0.20$ we find a curve enclosing a region
of ferromagnetism. For $\frac{kT}{t}\gtrsim 0.07$ these curves are closed and
lie away from the $\frac{t}{U}=0$ axis. Thus, at given density $n_{\text{h}}$
and temperature $T_{\text{C}}$, one has to go to finite $U$ to find a
transition. In other words: allowing for excitations in the spin background
enhances the ferromagnetic behavior. Furthermore, curves are generally not
enclosed by all contours at lower temperatures. This would imply that, at given
$n_{\text{h}}$ and $\frac{t}{U}$, one would find a para-ferro transition when
lowering the temperature from a region of high temperature, but also when
letting it increase from zero. This reentering of a paramagnetic phase at low
temperatures does not seem to be physical. It is in fact an artefact of this
method, due to convergence problems at very low temperatures. One can
understand this by looking at the expression~(\ref{invsushf}). If the
highest-order term becomes of order one, the series is clearly too short and
does not converge properly. This means that the results become unreliable for
$\frac{t}{U}\gtrsim\frac{kT}{2t}$ in the case of the square lattice, and
$\frac{t}{U}\gtrsim\frac{kT}{4t}$ on the simple cubic lattice. For a few curves
we have indicated this by a dashed line. As the approximations are better for
higher temperatures, we assume that the actual curve at $T_{\text{C}}=0$ (for
which we can only perform a calculation at infinite~$U$) should enclose all
curves shown.

In Figs.~\ref{square}b and~\ref{sc}b, we show Curie temperatures in contours of
fixed~$\frac{t}{U}$. Again we see the non-physical behavior of curves being
closed at the low-temperature side, for almost all values of $\frac{t}{U}$.
Fig.~\ref{n.09} shows Curie temperatures at fixed $n_{\text{h}}=0.09$, for the
simple cubic lattice, indicated by the dotted-dashed lines in Figs.~\ref{sc}a
and~\ref{sc}b. The dotted line in Fig.~\ref{n.09} indicates the region where
the series expansion becomes unreliable, according to the arguments presented
above.

There is one other point we want to mention here. As we have stated in the
introduction, we have also constructed the staggered susceptibility by
replacing the magnetic field $h$ by a staggered field $h_{\text{s}}$. Although
it is much more complicated to calculate the high-temperature expansions for
that case, as the number of terms involved increases significantly, it is not
difficult to obtain expressions for the staggered susceptibility, both at half
filling and in the one-hole approximation, for $h_{\text{s}} = 0$.
Thus, one may think that it is possible to obtain similar results for the
transition between a paramagnetic and an antiferromagnetic state, and conclude
which transition occurs first. When putting the inverse staggered
susceptibility at half filling (the equivalent of~(\ref{invsushf}) for the
antiferromagnetic system) to zero, one finds solutions for all values of the
parameter $\beta U$. This means that the staggered susceptibility of the
half-filled system diverges at a finite temperature. Apparently, the
para-antiferro transition is driven by the background itself, and may be
disturbed by a finite hole density. In our formulation, however, it is the
holes that drive the system into an ordered state, and the background only
indirectly contributes to the transition via its interaction with the holes.
This formulation is clearly not suitable to describe the transition to an
antiferromagnetic state. Therefore we only briefly indicate what we expect for
the para-antiferro transition.

In Fig.~\ref{af} we plot N\'eel temperatures for the simple cubic lattice at
half filling, in approximations to different orders in the parameter $\beta t$.
We see that the convergence of the series expansion is very good for large~$U$.
A transition from a paramagnetic to an antiferromagnetic phase is expected for
all values of $U$. It is at $T_{\text{N}}=0$ for infinite~$U$, and at
increasing temperatures with decreasing~$U$. For finite hole densities we
expect the transition to occur at lower temperatures, and at some point cross
the para-ferro transition.

\section{Discussion and conclusions}
\label{Discuss}
We have calculated Curie temperatures for the large-$U$ Hubbard model on the
square and simple cubic lattices, by means of a new extrapolation method to
extract information on low-temperature behavior from high-temperature series
expansions. We find a region of ferromagnetic behavior in the magnetic-phase
diagram, near half filling.

Comparing previous results for the simple cubic lattice, as depicted in
Fig.~\ref{contours}, to our current results, shown in Fig.~\ref{sc}, we see
that we now find a Curie temperature that is an order of magnitude smaller than
before. Furthermore, as we have checked in the case of infinite~$U$, subsequent
approximations in the current method do give consistent results, instead of
alternatingly producing Curie temperatures or not. These convergence problems
in the primitive series expansions are likely due to the fermi degeneracy of
the electron gas. At $\beta t \approx 1$, the wavelength of the electrons
becomes equal to the lattice distance, causing this degeneracy and divergences
to be present. When applying a straightforward extrapolation technique, one can
not account for this degeneracy, leading to results that are erroneous for
$\beta t \gtrsim 1$. In our approximation, using a density of states for holes,
we take the fermi degeneracy into account, and therefore we are able to proceed
to lower temperatures. We are confident that our present results do not suffer
from the above-mentioned convergence problems.

As we show in Fig.~\ref{curie2infU}, the difference between approximations to
different orders in the parameter $\beta t$ is rather small, and adding the
interaction also does not change the result considerably. Thus we believe the
eighth-order non-interacting hole approximation to be sufficient to describe
the qualitative behavior, and to obtain a good indication for numerical values.
We may add that, as a check, we have compared the free energy from calculations
by this method to results following directly from the series expansions, at
$\beta t\lesssim 0.5$, where the expansions are almost exact, and that these
results agree very well.

Our method works only for large~$U$, low hole density ($n_{\text{h}}\lesssim
0.2$), and, depending on the value of $U$, sufficiently high temperature. This
is clear from Figs.~\ref{square}--\ref{n.09}, where we see that the results are
unreliable for $\frac{t}{U}\gtrsim\frac{kT}{4t}$. We believe, however, that our
method gives a correct description for the tendencies in the half-filled system
at infinite~$U$, and for the qualitative behavior up to $n_{\text{h}}\approx
0.2$.

There are, however, some important limitations to this method, due to which we
are not able to predict a ferromagnetic state with certainty.

As we know from a theorem by Ghosh\cite{ghosh}, similar to the Mermin-Wagner
theorem for the Heisenberg model, the Hubbard model does not have long-range
ordering in two dimensions for finite temperatures. Thus, we must expect a
ferromagnetic phase in the two-dimensional case to be of the
Kosterlitz-Thouless type. Our method is essentially based on short-range
information from the high-temperature expansion (which is obtained via
calculations on small systems). It gives similar results  for the square and
the simple cubic lattices, as can be seen in Figs.~\ref{square} and~\ref{sc},
and we can not distinguish between different kinds of phases occurring.

Also, the method currently fails to describe the case of a para-antiferro
transition, due to the fact that a divergent background is not treated
correctly. We can therefore calculate only possible second-order phase
transitions between a paramagnetic and a ferromagnetic phase, for the case of
finite hole density. At half filling, we do find a finite N\'eel temperature
for any finite value of the parameter $\beta U$ (see Fig.~\ref{af}). This
implies that, near half filling, there is a transition from a paramagnetic to
an antiferromagnetic state at a higher temperature than the calculated
para-ferro transition. Thus, the para-ferro transition can not occur, and one
must study the antiferro-ferro transition to determine the ground-state
behavior.

Finally, due to the thermodynamic approach in which all possible states are
taken into account, our method can not distinguish special states that may
start to dominate  the system at low temperatures. Such states, if any, are not
recognized by the high-temperature expansion. An example of this is the fact
that it fails to reflect the influence of $m=1$ states in an $m=0$ system.
\vskip \baselineskip

We can compare our results to the work of Putikka et al.\cite{putikka}, who
calculate series expansions similar to those used by us, for the related
$t$--$J$ model, and extrapolate to low temperatures by means of Pad\' e
approximants. For $J>0$, in the limit of small $J$, the $t$--$J$ model is
equivalent to the large-$U$ Hubbard model. They find a region of weak
ferromagnetism (i.e. the spins are not fully aligned) for small positive $J$,
at hole density $n_{\text{h}} < 0.28 \pm 0.05$, which is in good agreement with
our results.

It is also encouraging to note that some of our results are in reasonable
quantitative agreement with results using an approximation of an entirely
different nature. By means of the slave-boson mean-field approach (at $T=0$),
Denteneer and Blaauboer\cite{denteneer} find a critical hole density
$n_{\text{h}}^{\text{c}}=1/3$ for ferromagnetism to occur at $U=\infty$, in
agreement with the values 0.27--0.29 found here (see Fig.~\ref{curie2infU}).
They also find that the value of $U/t$ above which ferromagnetism can occur, is
$U/t = 20$ (at $n_{\text{h}} = 0.17$), whereas one may extrapolate the results
of our Fig.~\ref{square}a to $T=0$ to find $U/t = 15$ (at $n_{\text{h}} =
0.15$).

Von der Linden and Edwards\cite{linden} use a variational approach to find a
ferromagnetic region in the $T=0$ phase diagram of the square-lattice Hubbard
model. They rigorously conclude that the state of complete spin alignment is
unstable when $n_{\text{h}} > 0.29$, for all $U$, and when $U/t < 42$, for all
$n_{\text{h}}$. The latter value is significantly higher than the value above
which we find ferromagnetism, but we assume that that is due to the fact that
they consider only strong ferromagnetism (full alignment of the spins), whereas
our method may also include weak ferromagnetism.

Also the results of Barbieri et al.\cite{manyholes}, who consider systems with
a large (but finite) number of holes, support the existence of ferromagnetic
behavior.

A final comparison that can be made is for the relation between the N\' eel
temperature and $U/t$ in the half-filled system. From Fig.~\ref{af} one can
calculate that the para-antiferro transition occurs for $kT_{\text{N}} \approx
3.85 t/U$. The large-$U$ Hubbard model at half filling is known to be
equivalent to an antiferromagnetic Heisenberg spin model, for which estimates
of the values of the critical temperature are given in
Ref.\onlinecite{rushbrooke}. According to the results mentioned there, the
relation would be $kT_{\text{N}} \approx 3.80 t/U$, which is in very good
agreement.
\vskip \baselineskip

This research was supported by the Stichting voor Fundamenteel Onderzoek der
Materie (FOM), which is financially supported by the Nederlandse Organisatie
voor Wetenschappelijk Onderzoek (NWO).

\section*{Appendix A: Enumeration of paths}
\setcounter{equation}{0}
\renewcommand{\theequation}{A.\arabic{equation}}
In this appendix we describe an efficient way to calculate the moments of the
density of states, for the case of infinite~$U$, by which we have calculated 22
of these moments for the square lattice, and 16 for the simple cubic lattice.
We start from Eq.~(\ref{rhom}), which we expand in terms of the parameter
$\beta t$:
\begin{equation} \label{Amom}
 \frac{Z_1^{\text{h}}}{N} = \sum_{n=0}^{\infty}(\beta
t)^n\frac{(-1)^n}{n!}M_n(m)
\end{equation}
with the moments of the density of states defined as
\begin{equation}
 M_n(m) \equiv \int d\varepsilon\bar{\rho}(\varepsilon,m)\varepsilon^n
\end{equation}
We can write the partition function for one hole according to its definition
(Cf.~(\ref{cpfholes})) also as
\begin{eqnarray} \label{Z1h}
 Z_1^{\text{h}} & \equiv & e^{(N-1)\beta\epsilon_{\text{hf}}}Z_{N-1} \\
 & = &
\!\left(\!\!\begin{array}{c
}N-1\\N_{\uparrow}(m)\end{array%
}\!\!\right)^{-1}\!\!\!\!\!\sum_{|i,\alpha_i(m)\rangle}\!\!\!\!\langle
i,\alpha_i(m)|e^{-\beta{\cal H}_{\text{kin}}}|i,\alpha_i(m)\rangle\!
\end{eqnarray}
where the summation is over all states $|i,\alpha_i(m)\rangle$ with a hole at
site $i$ and with a spin background $\alpha_i(m)$ such that the magnetization
per spin is indeed $m$. $N_{\uparrow}$ denotes the number of electrons with
spin up, which depends on $m$, and the factor
$\left(\!\begin{array}{c}N-1\\N_{\uparrow}(m)\end{array}\!\right)$ is the total
number of possible background configurations given the location of the hole,
which accounts for the spin degrees of freedom. In the thermodynamic limit,
this factor is exactly equal to the exponential factor in~(\ref{Z1h}), as one
easily checks by applying Stirling's formula for the binomial, and
with~(\ref{epsilon_hf}) for $\epsilon_{\text{hf}}$. The summation over $i$
gives a trivial (translational) factor $N$, and we can expand the exponential
in powers of $\beta t$ to obtain
\begin{equation} \label{Apat}
\frac{Z_1^{\text{h}}}{N}\left(\!\begin{array}{c
}N-1\\N_{\uparrow}(m)\end{array}\!\right) =
\sum_{|\alpha(m)\rangle}\sum_n\frac{(-1)^n}{n!}A_n(\alpha(m))(\beta t)^n
\end{equation}
where
\begin{equation}
 A_n(\alpha(m)) = \langle\alpha(m)|\left(\frac{{\cal
H}_{\text{kin}}}{t}\right)^n|\alpha(m)\rangle
\end{equation}
is the number of walks of length $n$ in the configuration space, that restore
the spin background $\alpha(m)$ to its original state. Comparing~(\ref{Amom})
and~(\ref{Apat}) we see that
\begin{equation}
 M_n(m) =
\left(\!\begin{array}{c}N-1\\N_{\uparrow}(m)\end{array%
}\!\right)^{-1}\sum_{|\alpha(m)\rangle}A_n(\alpha(m))
\end{equation}
Thus $M_n(m)$ is precisely the sum over all possible closed walks $w_n$ of
length $n$, summing the fraction of spin backgrounds that is restored by $w_n$.
Such a walk induces a permutation ${\cal P}(w_n)$ of the background spins,
which can be written as a product of disjunct cyclic permutations ${\cal
P}_i(w_n)$ with length $\left|{\cal P}_i(w_n)\right|>1$. In order to restore
the spin background $\alpha(m)$, the direction of the spin on each site must
remain unchanged, when applying ${\cal P}_i(w_n)$. Thus, all spins that are
interchanged by this permutation must point in the same direction. As the
number of spins involved is negligible compared to the total number of spins,
we may approximate that the probability to find an individual spin pointing up
or down is given by $\frac{1+m}{2}$ and $\frac{1-m}{2}$, respectively. Hence
the fraction of backgrounds in which the alignment of the spins remains
unchanged under the permutation ${\cal P}_i(w_n)$ is
$\left(\frac{1+m}{2}\right)^l+\left(\frac{1-m}{2}\right)^l$, where
$l=\left|{\cal P}_i(w_n)\right|$ is the number of spins involved in the
permutation. Thus, we can calculate $M_n$ as
\begin{equation} \label{Mmom}
 M_n(m) = \sum_{w_n}\prod_i\left(\!\left(\frac{1+m}{2}\right)^{|{\cal
P}_i(w_n)|}\!\!\!+\left(\frac{1-m}{2}\right)^{|{\cal P}_i(w_n)|}\right)
\end{equation}
For the actual evaluation of this expression we proved an elegant theorem that
enables us to significantly extend earlier calculations of the moments to
$n=22$. Defining a {\em retracing sequence\/} as two subsequent steps of the
hole in opposite directions (thus after two steps the hole is back in its
previous position; note that the last and first steps of a closed walk are
considered to be subsequent as well), one can make a distinction between {\em
reducible\/} and {\em irreducible\/} closed walks: an irreducible walk does not
contain any retracing sequence, whereas a reducible walk does. A reducible walk
can be made irreducible by repeatedly removing its retracing sequences; the
result is called the irreducible part of the walk. Note that a retracing
sequence does not permute spins, so the irreducible part of a walk induces the
same permutation of the spins as the walk itself. Thus, it is sufficient to
study only irreducible walks if one knows of how many reducible walks of a
given length it is the irreducible part. We proved the following formula: the
number of closed walks of length $l + 2n$ on a hypercubical lattice with
coordination number $z$, that have a given irreducible part of length $l>0$, is
\begin{equation}
 N_{\text{ir}}(l,n) = (z-1)^n\left(\begin{array}{c}l+2n\\n\end{array}\right)
\end{equation}
This greatly facilitates the calculation of~(\ref{Mmom}).

\section*{Appendix B: Inverse susceptibility in the interacting-hole
approximation}
\setcounter{equation}{0}
\renewcommand{\theequation}{B.\arabic{equation}}
In this appendix we give the formula for the inverse susceptibility in the
interacting-hole approximation, using the theory given in Sec.~\ref{Interact}.
We start from Eq.~(\ref{deffreeen}), which has to be differentiated with
respect to $m$ in order to get the equivalent of~(\ref{field}),
with~(\ref{Zgrint}) for $\ln Z_{\text{gr}}$:
\begin{eqnarray}
 \beta h = \beta h_{\text{hf}} & + &
n_{\text{h}}\frac{\partial\beta\mu_{\text{h}}}{n_{\text{s}}\partial m}
   -  \int
d\varepsilon\frac{\partial\bar{\rho}(\varepsilon)}{n_{\text{s}}\partial
m}\ln(1+e^{-\beta\tilde{\varepsilon}}) + \int
d\varepsilon\bar{\rho%
}(\varepsilon)n(\tilde{\varepsilon})\frac{\partial\beta\tilde{\varepsilon}}
{n_{\text{s}}\partial m}
  \nonumber \\ & - & \beta t\int d\varepsilon\int
d\varepsilon^\prime\frac{\partial\bar{\rho}(\varepsilon)}{n_{\text{s}}\partial
m}n(\tilde{\varepsilon})\bar{\rho%
}(\varepsilon^\prime)n(\tilde{\varepsilon%
}^\prime)\phi(\varepsilon,\varepsilon^\prime)
  \nonumber \\ & - & \beta t\int d\varepsilon\int
d\varepsilon^\prime\bar{\rho}(\varepsilon)\frac{\partial
n(\tilde{\varepsilon})}{n_{\text{s}}\partial
m}\bar{\rho}(\varepsilon^\prime)n(\tilde{\varepsilon%
}^\prime)\phi(\varepsilon,\varepsilon^\prime)
 \nonumber \\ & - & \frac{\beta t}{2}\int d\varepsilon\int
d\varepsilon^\prime\bar{\rho}(\varepsilon)n(\tilde{\varepsilon%
})\bar{\rho}(\varepsilon^\prime)n(\tilde{\varepsilon%
}^\prime)\frac{\partial\phi(\varepsilon,\varepsilon^\prime)}
{n_{\text{s}}\partial m}
\label{fieldint} \end{eqnarray}
where $n_{\text{h}}$ is given by
\begin{equation} \label{holdenint}
 n_{\text{h}}  =  -\int
d\varepsilon\bar{\rho}(\varepsilon)n(\tilde{\varepsilon%
})\frac{\partial\beta\tilde{\varepsilon}}{\partial\beta\mu_{\text{h}}}
  + \beta t\int d\varepsilon\int
d\varepsilon^\prime\bar{\rho}(\varepsilon)n(\tilde{\varepsilon%
})\bar{\rho}(\varepsilon^\prime)\frac{\partial n(\tilde{\varepsilon%
}^\prime)}{\partial\beta\mu_{\text{h}}}\phi(\varepsilon,\varepsilon^\prime)
\end{equation}
This may look awkward, but if we look at the derivatives of
$\tilde{\varepsilon}$ (see Eq.~(\ref{epstilde})) we see that many of these
terms cancel. Let us first look at the expression~(\ref{holdenint}) for the
hole density. As we are working at fixed hole density, derivatives of the fermi
factor do not play a role in these equations, and they vanish. We need the
derivative of $\tilde{\varepsilon}$ with respect to $\beta\mu_{\text{h}}$,
\begin{equation}
 \frac{\partial\beta\tilde{\varepsilon}}{\partial\beta\mu_{\text{h}}} = -1 +
\beta t\int
d\varepsilon^\prime\bar{\rho%
}(\varepsilon^\prime)\phi(\varepsilon,\varepsilon^\prime)\frac{\partial
n(\tilde{\varepsilon}^\prime)}{\partial\beta\mu_{\text{h}}}
\end{equation}
and so we see that indeed there is a cancellation of terms, leaving us with the
relation
\begin{equation}
 n_{\text{h}} = \int d\varepsilon\bar{\rho}(\varepsilon)n(\tilde{\varepsilon})
\end{equation}
Then, we rewrite the expression for the magnetic field with
\begin{eqnarray}
 \frac{\partial\beta\tilde{\varepsilon}}{n_{\text{s}}\partial m}  =
-\frac{\partial\beta\mu_{\text{h}}}{n_{\text{s}}\partial m} & + & \beta t\int
d\varepsilon^\prime\frac{\partial\bar{\rho}(\varepsilon^\prime)}
{n_{\text{s}}\partial
m}n(\tilde{\varepsilon}^\prime)\phi(\varepsilon,\varepsilon^\prime)
   +  \beta t\int
d\varepsilon^\prime\bar{\rho}(\varepsilon^\prime)\frac{\partial
n(\tilde{\varepsilon}^\prime)}{n_{\text{s}}\partial
m}\phi(\varepsilon,\varepsilon^\prime)
   \nonumber \\ & + & \beta t\int
d\varepsilon^\prime\bar{\rho}(\varepsilon^\prime)n(\tilde{\varepsilon%
}^\prime)\frac{\partial\phi(\varepsilon,\varepsilon^\prime)}
{n_{\text{s}}\partial m}
\end{eqnarray}
Using this expression it is straightforward to check that~(\ref{fieldint})
reduces to
\begin{equation}
 \beta h  =  \beta h_{\text{hf}} - \int
d\varepsilon\frac{\partial\bar{\rho}(\varepsilon)}{n_{\text{s}}\partial
m}\ln(1+e^{-\beta\tilde{\varepsilon}})
   +  \frac{\beta t}{2}\int d\varepsilon\int
d\varepsilon^\prime\bar{\rho}(\varepsilon)n(\tilde{\varepsilon%
})\bar{\rho}(\varepsilon^\prime)n(\tilde{\varepsilon%
}^\prime)\frac{\partial\phi(\varepsilon,\varepsilon^\prime)}
{n_{\text{s}}\partial m}
\label{hint} \end{equation}
In order to derive the inverse susceptibility from this expression, we have to
take the derivative with respect to $n_{\text{s}}m$ again, and put $m=0$. For
reasons of symmetry it is easy to show that the first derivatives with respect
to $m$ of all functions appearing in the integrals vanish at $m=0$. Thus, in
the terms in~(\ref{hint})
we only have to consider the derivatives of the functions that have been
differentiated once already:
\begin{eqnarray}
 \beta N\chi_{\text{fm}}^{-1} = \left.\frac{\partial\beta
h_{\text{hf}}}{n_{\text{s}}\partial m}\right|_{m=0} & - & \int
d\varepsilon\left.\frac{\partial^2\bar{\rho}(\varepsilon)}
{n_{\text{s}}^2\partial m^2}\right|_{m=0}\ln(1+e^{-\beta\tilde{\varepsilon}})
  \nonumber \\ & + & \frac{\beta t}{2}\int d\varepsilon\int
d\varepsilon^\prime\bar{\rho}(\varepsilon)n(\tilde{\varepsilon%
})\bar{\rho}(\varepsilon^\prime)n(\tilde{\varepsilon%
}^\prime)\left.\frac{\partial^2\phi(\varepsilon,\varepsilon^\prime)}
{n_{\text{s}}^2\partial m^2}\right|_{m=0}
\end{eqnarray}
This can again be expressed in terms of $\rho(\varepsilon)$ (note that also
$\phi$ is being legendre transformed):
\begin{eqnarray}
  \beta N\chi_{\text{fm}}^{-1} =  \left.\left.\frac{\partial\beta
h_{\text{hf}}}{n_{\text{s}}\partial m}\right|_{m=0}\right(1 & - &
\left.\frac{\partial\beta h_{\text{hf}}}{n_{\text{s}}\partial
m}\right|_{m=0}\left[\int
d\varepsilon\frac{\partial^2\rho(\varepsilon)}{\partial(\beta
h)^2}\ln(1+e^{-\beta\tilde{\varepsilon}})
 \right. \nonumber \\ & + & \left.\left. \frac{\beta t}{2}\int d\varepsilon\int
d\varepsilon^\prime\rho(\varepsilon)n(\tilde{\varepsilon%
})\rho(\varepsilon^\prime)n(\tilde{\varepsilon%
}^\prime)\frac{\partial^2\phi(\varepsilon,\varepsilon^\prime)}
{\partial(\beta h)^2}\right]_{h=0}\right)
\end{eqnarray}
which is the modification of~(\ref{invsush}) for interacting holes.

\newpage \section*{Tables}
\begin{table}
\refstepcounter{table}
\label{moments}
\begin{center}\vspace*{-2ex}
\begin{tabular}{|r@{~~}|r@{}l|r@{~~}|}
 \hline
 $~~n$ & \multicolumn{2}{c|}{$m=0$} & \multicolumn{1}{c|}{$m=1$}  \\
 \hline
 0 &           1&     &            1 \\
 2 &           4&     &            4 \\
 4 &          30&     &           36 \\
 6 &         269&.5   &          400 \\
 8 &        2641&.75  &         4900 \\
10 &       27279&.94  &        63504 \\
12 &      291718&.97  &       853776 \\
14 &     3199250&.29  &     11778624 \\
16 &    35766660&.22  &    165636900 \\
18 &   405989247&.14  &   2363904400 \\
20 &  4665921461&.01  &  34134779536 \\
22 & ~~54182396281&.84~~  & ~~497634306624 \\
\hline
\end{tabular}
\end{center}
TABLE~\ref{moments}. The first 22
moments of $\bar{\rho}(\varepsilon)$ for $m=0$ and $m=1$, for the square
lattice (odd moments vanish).
\end{table}

\begin{table}
\refstepcounter{table}
\label{intcoef}
\begin{center}\vspace*{-2ex}
\begin{tabular}{|r@{~~}|r@{}l|c|}
 \hline
 $~~k$ & \multicolumn{2}{c|}{$m=0$} & ~~$m=1$~~  \\
 \hline
 0 &    0&                & 0 \\
 2 &    0&                & 0 \\
 4 &   -0&.333333333333   & 0 \\
 6 &   -0&.5875           & 0 \\
 8 &   -0&.424851190476   & 0 \\
10 &   -0&.170455970293   & 0 \\
12 & ~~-0&.041839059880~~ & 0 \\
\hline
\end{tabular}
\end{center}
TABLE~\ref{intcoef}. Values of
$\frac{1}{N}\left(U_2^{\text{int}}\right)_k$
for $m=0$ and $m=1$, for the square lattice with $U=\infty$.
\end{table}

\begin{table}
\refstepcounter{table}
\label{Umoments}
\begin{center}\vspace*{-2ex}
\begin{tabular}{|c|@{~~~}l|}
 \hline & \\[-2mm]
 $~~n~~$ & $\int d\varepsilon\rho(\varepsilon)\varepsilon^n$  \\
 & \\[-2mm] \hline & \\[-2mm]
 0 & 1 \\
 2 & $2\left(2 - \frac{2}{\beta U}\right)$ \\
 4 & $24\left(\frac{5}{4} - \frac{2}{\beta U} - \frac{12}{(\beta U)^2} +
\frac{3}{(\beta U)^3}\right)$ \\
 6 & $720\left(\frac{539}{1440} - \frac{59}{48\beta U} - \frac{93}{8(\beta
U)^2} - \frac{89}{6(\beta U)^3} + \frac{1}{2(\beta U)^4} + \frac{127}{2(\beta
U)^5}\right)$ \\
 8 & $40320\left(\frac{10567}{161280} - \frac{271}{576\beta U} -
\frac{1459}{320(\beta U)^2} - \frac{377}{32(\beta U)^3} + \frac{4531}{96(\beta
U)^4} + \frac{4043}{8(\beta U)^5} - \frac{28837}{8(\beta U)^6} +
\frac{78593}{8(\beta U)^7}\right)$~~~~~~~~~~~~~ \\[-2mm]
 & \\ \hline
\end{tabular}
\end{center}
TABLE~\ref{Umoments}. Moments of the density of states for the square lattice
(odd moments vanish), at $h=0$.
\end{table}

\newpage \section*{Figure captions}
\begin{figure}
\refstepcounter{figure}
\label{divergence}
FIG.~\ref{divergence}. The inverse ferromagnetic susceptibility as a function
of the parameter $\beta t$, for the Hubbard model on a simple cubic lattice,
with infinite $U$ and particle density $n=0.9$. Approximations up to order 2,
4, 6, 8 in $\beta t$ obtained by means of the cluster expansion method.
\end{figure}

\begin{figure}
\refstepcounter{figure}
\label{contours}
FIG.~\ref{contours}. N\'eel and Curie temperatures, as a function of the
particle density, for the Hubbard model on a simple cubic lattice, at constant
$t/U$.
\end{figure}

\begin{figure}
\refstepcounter{figure}
\label{density_of_states}
FIG.~\ref{density_of_states}a. The density of states for a paramagnetic system
on a square lattice, using up to the number of moments indicated.
\end{figure}

\begin{figure}
FIG.~\ref{density_of_states}b. The density of states in the ferromagnetic
regime on a square lattice. Exact result, and approximations using up to the
number of moments indicated.
\end{figure}

\begin{figure}
\refstepcounter{figure}
\label{curie2infU}
FIG.~\ref{curie2infU}. Curie temperatures (contours of zero inverse
ferromagnetic susceptibility)  for the square lattice at infinite~$U$. \\
(a) Non-interacting hole approximation, 8th order; \\
(b) Non-interacting hole approximation, 22nd order; \\
(c) Interacting-hole approximation, 22nd order.
\end{figure}

\begin{figure}
\refstepcounter{figure}
\label{square}
FIG.~\ref{square}. Magnetic-phase diagram for the square lattice.\\
(a) Contours of fixed Curie temperature, with $kT_{\text{C}}/t = 0.03$, 0.04,
\ldots, 0.19 (increment 0.01).\\
(b) Curie temperature at fixed $t/U = 0, 0.005$, \ldots, 0.055 (increment
0.005).
\end{figure}

\begin{figure}
\refstepcounter{figure}
\label{sc}
FIG.~\ref{sc}. Phase diagram for the simple cubic lattice.\\
(a) Contours of fixed Curie temperature, with $kT_{\text{C}}/t = 0.03$, 0.04,
\ldots, 0.14 (increment 0.01). \\
(b) Curie temperature at fixed $t/U = 0$, 0.002, \ldots, 0.022 (increment
0.002).
\end{figure}

\begin{figure}
\refstepcounter{figure}
\label{n.09}
FIG.~\ref{n.09}. Curie temperature for the simple cubic lattice, at
$n_{\text{h}} = 0.09$. The dashed part of the curve is unreliable, due to lack
of convergence (indicated by the dotted line).
\end{figure}

\begin{figure}
\refstepcounter{figure}
\label{af}
FIG.~\ref{af} N\'eel temperature for the simple cubic lattice at half filling.
Approximations to different orders in $\beta t$, as indicated.
\end{figure}

\end{document}